\documentclass[a4paper]{article}

\usepackage{INTERSPEECH2022}

\usepackage{comment}
\usepackage{enumitem}
\usepackage{diagbox}
\usepackage{multirow}
\usepackage{colortbl}
\usepackage{booktabs}
\usepackage{extarrows}
\usepackage{xspace}
\usepackage{xcolor}
\usepackage{tcolorbox}
\usepackage{soul}
\usepackage{hyperref}
\usepackage{microtype} 
\usepackage[noabbrev]{cleveref} 
\newlist{todolist}{itemize}{2}
\setlist[todolist]{label=$\square$, leftmargin=0.5mm}

\newenvironment{todo}{%
    \begin{tcolorbox}[colback=red!5!white,colframe=red!75!black,title=TODO]
    \begin{todolist}
}{%
    \end{todolist}
    \end{tcolorbox}
}

\newcommand{\fixme}[1]{\xspace\textcolor{red}{\hl{\textbf{#1}}}}

\newcommand{\MKCLEAN}{
    \renewcommand{\fixme}[1]{}
    
}
\MKCLEAN 

\definecolor{mtlcolor}{rgb}{1,0.918,0.773}
\newcommand{\hlo}[1]{{\sethlcolor{mtlcolor}\hl{#1}}}

\DeclareMathOperator*{\argmin}{arg\,min}

\def\lambdaTA{\lambda^{(t)}_{A}}
\def\lambdaTC{\lambda^{(t)}_{C}}
\def\lambdaIC{\lambda^{(i)}_{C}}

\title{
Hear No Evil:
Towards Adversarial Robustness\\
of Automatic Speech Recognition
via Multi-Task~Learning
}
\name{Nilaksh Das, Duen Horng Chau}
\address{Georgia Institute of Technology, USA}
\email{\{nilakshdas,polo\}@gatech.edu}

\begin{document}

\maketitle
%

\begin{abstract}
As automatic speech recognition (ASR) systems
are now being widely deployed in the wild,
the increasing threat of adversarial attacks
raises serious questions
about the security and reliability
of using such systems.
On the other hand,
multi-task~learning (MTL)
has shown success in training
models
that can resist 
adversarial attacks
in the computer vision domain.
In this work,
we investigate the impact of
performing such multi-task learning
on the adversarial robustness
of ASR models
in the speech domain.
We conduct extensive MTL experimentation
by combining semantically diverse tasks
such as accent classification and ASR,
and evaluate a wide range of adversarial settings.
Our thorough analysis reveals
that
performing MTL 
with semantically diverse tasks
consistently makes it harder
for an adversarial attack to succeed.
We also discuss in detail
the serious pitfalls
and their related remedies
that have a significant impact
on the robustness of MTL models.
Our proposed MTL approach shows 
considerable absolute improvements in 
adversarially targeted WER 
ranging from 17.25 up to 59.90 
compared to single-task learning baselines 
(attention decoder and CTC respectively).
Ours is the first in-depth study
that uncovers adversarial robustness gains
from multi-task learning for ASR.
\end{abstract}


\section{Introduction}

Automatic speech recognition (ASR) systems
have penetrated our daily lives
with real-world applications
such as
digital voice assistants, IVR
and news transcription.
These are increasingly relying
on deep learning methods
for their superior performance.
At the same time,
a mature body of
adversarial machine learning research
has exposed serious vulnerabilities
in these underlying methods~\cite{cisse2017houdini,iter2017generating,alzantot2018did,carlini2018audio,yuan2018commandersong,taori2019targeted,wang2020towards},
raising grave concerns
regarding the trustworthiness
of ASR applications.
There is an urgent need
to address these vulnerabilities
for restoring faith in using
ASR for safety-critical functions.
Our study aims to push the envelope
in this direction
with a meticulous approach.

An adversarial attack on an ASR model
allows the attacker
to introduce faint noise
to a speech sample
that can influence the model
into making an exact transcription
of the attacker's choosing.
Since this targeted adversarial scenario
is considered more threatening
for an ASR system as compared to
adding noise that leads to
some arbitrary prediction~\cite{alzantot2018did,taori2019targeted},
we focus on studying the characteristics
of ASR models that can thwart
\textit{targeted} adversarial attacks.
Research has shown that
adversarial examples
are manifestations
of non-robust features
learned by a deep learning model~\cite{ilyas2019adversarial}.
Hence, our objective is to
regularize the ASR model training paradigm
so as to learn robust features
that can resist such attacks.

\textbf{Multi-task learning (MTL)}
is one such approach that has shown
some success in this aspect
for computer vision tasks
by making the underlying models
more resilient to adversarial attacks~\cite{mao2020multitask,ghamizi2021adversarial}.
However, it is unclear whether such robustness
would transfer to the audio modality.
Moreover, hybrid ASR models
are often trained with semantically equivalent
tasks like CTC and attention decoding~\cite{watanabe2017hybrid}.
This raises interesting questions
about their inherent adversarial robustness.
In this work,
we aim to study the impact
of MTL
on the adversarial robustness
of ASR models,
and compare it to robustness
of single-task learning (STL).
Here, we consider MTL
as jointly training a shared feature encoder
with multiple losses from
diverse task heads.
Our expectation is that
MTL would induce the encoder
to learn a robust feature space that
is harder to attack~\cite{ilyas2019adversarial}.
We consider semantically equivalent
as well as semantically diverse tasks
for performing MTL.
We find that a combination
of both types of tasks
is necessary to most effectively
thwart adversarial attacks.
We also discuss serious pitfalls
related to inference
for MTL models
that have an adverse impact
on robustness.
Finally, we demonstrate remedies
for such pitfalls
that significantly make it harder
for an attacker to succeed.

\smallskip
\noindent \textbf{Contributions}
\begin{itemize}[label=$\bullet$,leftmargin=*,noitemsep,topsep=0pt]
    \item \textbf{First MTL Study of Adversarial Robustness for ASR.}
    To the best of our knowledge,
    this is the first work to uncover adversarial robustness gains
    from MTL for ASR models.

    \item \textbf{Extensive Evaluation.}
    We perform extensive experimentation
    with one of the most powerful adversarial attacks,
    and evaluate models trained with semantically equivalent
    as well as semantically diverse tasks
    across a wide range of training hyperparameters
    and strong adversarial settings.

    \item \textbf{Robustness of Hybrid ASR Inference.}
    Our study exposes
    the extreme vulnerability of using CTC head for inference
    in hybrid CTC/attention models; while showing that MTL training
    with CTC and attention loss improves resiliency to adversarial attacks
    if CTC head is dropped during inference.

    \item \textbf{Robust ASR with Semantically Diverse Tasks.}
    Our thorough analysis reveals that
    performing MTL
    with semantically diverse tasks
    such as combining ASR with accent classification
    makes it most difficult for an attacker
    to induce a maliciously targeted prediction.
    Our MTL approach shows
    considerable absolute improvements in
    adversarially targeted WER
    ranging from 17.25 up to 59.90
    compared to STL baselines
    (attention decoder and CTC respectively).
\end{itemize}
\section{Related Works}

Several adversarial attacks
have been proposed
for maliciously influencing 
ASR models
by leveraging model gradients
to optimize
a faint perturbation
to the input speech~\cite{cisse2017houdini,iter2017generating,carlini2018audio,yuan2018commandersong,wang2020towards}.
Many such gradient-based 
adversarial techniques
can be formulated as variations
to the projected gradient descent (PGD) method~\cite{madry2018towards},
which is one of the strongest digital perturbation attacks
proposed in the adversarial ML literature.
In this work,
we experiment with the
\textit{targeted} PGD attack,
as this attack scenario
is considered more threatening
for ASR systems~\cite{alzantot2018did,taori2019targeted}.

Defenses proposed to evade adversarial attacks on ASR models
mostly employ input preprocessing~\cite{das2018adagio,hussain2021waveguard,sreeram2021perceptual,zelasko2021adversarial}
that places an undue burden on inference-time computation.
Adversarial training has
also shown some success 
in improving adversarial robustness~\cite{madry2018towards,yang2020characterizing}.
However, these methods are extremely computationally expensive.
MTL
with a shared backbone
has the potential
to provide a reasonable middle ground
as it is much less computationally expensive
than adversarial training,
and does not introduce any inference-time load.
Many studies have indeed looked at MTL
in the context of ASR robustness~\cite{watanabe2017hybrid,jain2018improved,adi2019reverse,das2021best,li2021adversarial,tang2021general}.
However, the bulk of such works
focus mostly on improvement in benign performance (when no attack is performed)
or robustness to arbitrary background noise.
Ours is the first work
to study the impact of MTL
on ASR models
specifically in the context of 
adversarial attacks.

\section{Approach}

\setlength{\abovedisplayskip}{5pt}
\setlength{\belowdisplayskip}{5pt}
\setlength{\abovedisplayshortskip}{0pt}
\setlength{\belowdisplayshortskip}{0pt}


%
%

\subsection{ASR with Multi-Task Learning}
\label{sec:approach_mtl}
\vspace{-0.4em}

In this work,
we study the impact of MTL
on adversarial robustness
of ASR models
by jointly training
a shared feature encoder $\phi$.
The encoder takes a speech sample
$x$ as input,
and outputs $\phi(x)$,
which can be considered
as a latent sequence embedding
in a shared feature space.
The feature embedding is
then passed to various task
heads with corresponding losses.
We consider semantically diverse tasks
in addition to semantically equivalent tasks
for performing MTL.
The semantically equivalent task heads
for ASR are:
(1) CTC and (2) attention decoder.
We also jointly train a discriminator
for a semantically diverse
task such as accent classification.

For the ASR task, we denote
the ground-truth transcription as $\bar{y}$,
and the CTC loss~\cite{graves2006connectionist}
and decoder attention loss~\cite{bahdanau2014neural}
as $\mathcal{L}_\text{CTC}$ and $\mathcal{L}_\text{DEC}$
respectively.
We compute the loss for ASR as:
\begin{align}
    \mathcal{L}_\text{ASR}(x, \bar{y}) = \lambdaTC \mathcal{L}_\text{CTC}(x, \bar{y}) + (1 - \lambdaTC) \mathcal{L}_\text{DEC}(x, \bar{y}) \label{eq:train_asr}
\end{align}

Consequently,
for performing joint ASR inference,
we denote the CTC and decoder
scoring functions~\cite{watanabe2017hybrid}
as $f_\text{CTC}$ and $f_\text{DEC}$
respectively.
Finally,
we determine the predicted output transcription $\hat{y}$
as follows:
\begin{align}
    \hat{y} = \lambdaIC f_\text{CTC}\big(\phi(x)\big) + (1 - \lambdaIC) f_\text{DEC}\big(\phi(x)\big) \label{eq:infer_asr}
\end{align}

Note here that $\lambdaTC$ and $\lambdaIC$
are training and inference weights respectively.
Generally, we follow that $\lambdaIC=\lambdaTC$
in our experiments while performing inference,
unless otherwise specified.
Correspondingly,
setting $\lambdaIC${}$=$1.0
allows us to use only the
trained CTC head for inference,
and vice-versa for the trained decoder head
by setting $\lambdaIC${}$=$0.0.

For accent classification,
we are given an accent label $\bar{z}$
that is to be predicted for a speech sample $x$.
Besides being semantically diverse,
accent classification is also
functionally distinct
as it is a \textit{sequence-to-label} task,
compared to the \textit{sequence-to-sequence} task for ASR.
Hence, denoting the cross-entropy loss
for the discriminator head
as $\mathcal{L}_\text{DIS}$,
we compute the full MTL loss
for jointly training the model as:
\begin{align}
    \mathcal{L}_\text{MTL}(x, \bar{y}, \bar{z}) = \lambdaTA \mathcal{L}_\text{ASR}(x, \bar{y}) + (1 - \lambdaTA) \mathcal{L}_\text{DIS}(x, \bar{z}) \label{eq:train_mtl}
\end{align}

From \Cref{eq:train_asr,eq:train_mtl},
we can see that modulating
the $\lambdaTA$ and $\lambdaTC$ weights
allows us to independently modulate
the effect of various heads during training,
\textit{e.g.}, setting $\lambdaTA${}$=$1.0 and $\lambdaTC${}$=$1.0
allows us to train using only the CTC head.
Conversely, we can train only the decoder head
by setting $\lambdaTA${}$=$1.0 and $\lambdaTC${}$=$0.0.
These can also be considered as the single-task learning (STL) baselines.
With $\lambdaTA < 1.0$,
we can train the model
with the discriminator head included.
We perform extensive experiments
across a range of these hyperparameters
to study the impact of MTL
on ASR robustness to attacks.

\subsection{Adversarial Attack on ASR with PGD}
\label{sec:approach_adv}
\vspace{-0.4em}

An adversarial attack on ASR introduces
an inconspicuous and negligible perturbation $\delta$
to a speech sample that confuses the ASR model
into making an incorrect prediction.
In this work, we focus on
the projected gradient descent (PGD) attack~\cite{madry2018towards}.
Specifically,
we consider the \textit{targeted} PGD attack,
as it is considered more malicious and threatening
for ASR systems~\cite{alzantot2018did,taori2019targeted}.
Given, a target transcription $\tilde{y}$,
the targeted attack aims to minimize
the following inference loss function
so as to force the ASR model into
making a prediction of the attacker's choosing:
\begin{align}
    \mathcal{L}_\text{ADV}(x, \tilde{y}) = \lambdaIC \mathcal{L}_\text{CTC}(x, \tilde{y}) + (1 - \lambdaIC) \mathcal{L}_\text{DEC}(x, \tilde{y})
\end{align}

The PGD attack is an iterative attack
that consists of two main stages.
The first stage is the perturbation stage,
wherein a small perturbation of step size $\alpha$
is computed in the direction of the gradient
of $\mathcal{L}_\text{ADV}$
with respect to the sample
from the previous iteration.
This perturbation having a magnitude of $\alpha$
is added to the sample.
The second stage is the projection stage
that ensures that the perturbed sample
remains within an $\epsilon$-ball
of the original input.
%
This is also called the $L2$ threat model,
as it uses the $L2$ norm
for limiting the perturbation.
Hence, the targeted PGD attack
optimizes the perturbation $\delta$ as:
\begin{align}
    x_\text{ADV} = \argmin_\delta \mathcal{L}_\text{ADV}(x + \delta, \tilde{y}), \text{   s.t.   } \|\delta\|_2 \leq \epsilon
\end{align}

The perturbation and projection stages
are performed iteratively,
and the computational cost to the attacker
increases with increasing number of iterations.
In order to isolate and study
the impact of our MTL training
on adversarial robustness,
we perform greedy ASR inference
while implementing the attack.
Since the adversarial objective is to
induce a maliciously targeted prediction
as opposed to untargeted arbitrary predictions,
we analyze the MTL robustness
to specifically evade such targeted attacks.
Hence, we examine the
adversarially targeted word error rate
(abbreviated as \textbf{AdvTWER} hereon)
in this work,
which reports the word error rate
of the prediction $\hat{y}$
with respect to the
adversarial target transcription $\tilde{y}$,
\textit{i.e.,} a higher AdvTWER implies
that the model is more robust in evading the attack.
Conversely, a lower AdvTWER
means that
the attack was more successful.
We report the
adversarially targeted word error rate
as we find it to be more compelling
in studying the MTL robustness.
Correspondingly, we also observe
equivalent robustness trends
with respect to the benign word error rate.

\section{Experiment Setup}

\begin{table*}[!t]
\centering
\caption{AdvTWER ($\uparrow$ is more robust) with $\lambdaTA${}$=${}$1.0$. Training with CTC and dropping CTC for inference ($\lambdaIC${}$=${}$0.0$) is more robust.}
\scalebox{0.9}{
\begin{tabular}{l|ccccr|ccc} 
\toprule
& \multicolumn{5}{c|}{$\lambdaIC=\lambdaTC$} & \multicolumn{3}{c}{$\lambdaIC=0.0$} \\[0.5em]
& \multicolumn{5}{c|}{\begin{tabular}[l]{@{\hspace{-6pt}}c@{}}\textcolor{gray}{\scriptsize STL}\vspace{-0.4em}\\\textcolor{gray}{\scriptsize (Decoder)}\end{tabular} $\xlongleftrightarrow{\hspace{1.8em}\text{MTL with joint inference}\hspace{1.8em}}$ \begin{tabular}[c]{@{}c@{}}\textcolor{gray}{\scriptsize STL}\vspace{-0.4em}\\\textcolor{gray}{\scriptsize (CTC)}\end{tabular}} & \multicolumn{3}{c}{\scriptsize \begin{tabular}[c]{@{}c@{}}MTL\\with Decoder inference\end{tabular}} \\
\multicolumn{1}{r|}{$\lambdaTC \xrightarrow{}$} & \multicolumn{1}{c}{$0.0$} & \multicolumn{1}{c}{$0.3$} & \multicolumn{1}{c}{$0.5$} & \multicolumn{1}{c}{$0.7$} & \multicolumn{1}{c|}{$1.0$} & \multicolumn{1}{c}{$0.3$} & \multicolumn{1}{c}{$0.5$} & \multicolumn{1}{c}{$0.7$} \\
\midrule
PGD-500 & \textcolor{gray}{93.35} & 59.87 & 61.55 & 76.64 & \textcolor{gray}{37.53} & {\cellcolor{mtlcolor}}{96.80} & {\cellcolor{mtlcolor}}{\textbf{99.95}} & {\cellcolor{mtlcolor}}{97.06} \\
PGD-1000 & \textcolor{gray}{72.31} & 35.21 & 37.49 & 53.91 & \textcolor{gray}{15.19} & {\cellcolor{mtlcolor}}{79.36} & {\cellcolor{mtlcolor}}{81.29} & {\cellcolor{mtlcolor}}{\textbf{82.45}} \\
PGD-1500 & \textcolor{gray}{57.57} & 23.94 & 26.09 & 42.44 & \textcolor{gray}{8.99} & {\cellcolor{mtlcolor}}{67.16} & {\cellcolor{mtlcolor}}{68.45} & {\cellcolor{mtlcolor}}{\textbf{69.57}} \\
PGD-2000 & \textcolor{gray}{49.79} & 19.22 & 20.40 & 36.99 & \textcolor{gray}{7.14} & {\cellcolor{mtlcolor}}{60.15} & {\cellcolor{mtlcolor}}{\textbf{63.29}} & {\cellcolor{mtlcolor}}{60.56} \\
\bottomrule
\multicolumn{9}{l}{\scriptsize \textcolor{gray}{gray} = STL; \hlo{highlighted} = MTL $>$ STL; \textbf{bold} = highest in row}
\end{tabular}
} 
\vspace{-1em}
\label{tab:dec_ctc}
\end{table*}

\subsection{Data}
\vspace{-0.4em}

We use annotated speech data
from Mozilla's Common Voice dataset~\cite{mozillacommonvoice}
for our experiments.
Common Voice consists of
naturally spoken human speech
in a variety of languages.
%
The dataset also includes
demographic metadata like age, sex and accent
that is self-reported by speakers,
and the speech is validated by annotators.
Specifically,
we use the English language speech
and extract
accent-labeled speech samples
for US and Indian accents,
which are among the most
abundantly available
accented speech in the dataset.
Using the splits
as defined by the dataset,
we get $\sim$260K samples for training
%
and $\sim$1.2K samples for validation.
Finally, we report the performance metrics
on $\sim$1K samples obtained
from the test split.

\subsection{Model Architecture and Training}
\vspace{-0.4em}

We leverage a pre-trained hybrid \mbox{CTC-attention} model
that is publicly available~\cite{kamo2021pretrained},
and fine-tune it using multi-task learning.
The model consists of
a conformer-based encoder~\cite{gulati2020conformer}
that is shared by multiple task heads.
The encoder has 12 conformer blocks,
each with 8 attention heads and
a hidden unit size of 2048.
The output size of the encoder is 512,
which is consequently the size of
the sequence embeddings
from the shared feature space.
The MTL model employs
a CTC head and a decoder head
for ASR.
The decoder has 6 transformer blocks,
with 8 attention heads and
a hidden unit size of 2048.
The ASR heads output scores
for 5000 subword units.
For accent classification,
we use a dense feed-forward discriminator
that passes the mean sequence embedding
through 5 fully connected layers,
and finally outputs class labels
corresponding to the accents.
The MTL models are implemented
using the ESPnet2 toolkit~\cite{watanabe2018espnet}.
We train several MTL models
by extensively modulating
the $\lambdaTA$ and $\lambdaTC$ weights.
Each MTL model is trained
with a learning rate of $10^{-3}$
for 30 epochs,
and we pick the model
with the lowest validation loss
from all the epochs.
%
%

\subsection{Adversarial Setting}
\label{sec:setup_adversarial}
\vspace{-0.4em}

For the targeted PGD attack,
we generate a fixed set
of adversarial transcriptions
having varying lengths.
For each sample,
the target transcription $\tilde{y}$
is chosen in a deterministic manner by
finding the adversarial transcription
from the fixed set
having the closest length
to the original transcription $\bar{y}$.
To ensure there is minimal word overlap
between the original transcriptions
and the adversarial transcriptions,
we use a lorem-ipsum generator
for generating the adversarial transcriptions.
We leverage the targeted PGD attack
implemented using the
Adversarial Robustness Toolbox~\cite{art2018}
for our experiments.
While attacking the ASR model,
we focus solely on the MTL robustness
by performing greedy inference
for determining the gradients,
and no additional LMs are used.
The targeted PGD attack
is performed using an $L2$ threat model
in a white-box attack for inference.
We use an
extremely strong perturbation limit
of $\epsilon${}$=$2.0,
beyond which we observed that the perturbed samples
would no longer remain within natural constraints.
As the attack's computational cost
increases with increasing number of steps,
we report the AdvTWER metric
for performing multiple attack steps,
up to as high as 2000 steps.
We use a step size $\alpha${}$=$0.05,
allowing the attack to make fine-grained
perturbations at each step.

\section{Results}

Our extensive experimentation
reveals that increasing the proportion
of multi-task learning (MTL)
while training
significantly improves the
model's resiliency to adversarial attacks.
Before analysing the adversarial performance
of the models,
we first briefly discuss
the benign performance,
\textit{i.e.}, when no attack is performed.
On the test set,
the single-task learning (STL)
baselines have a
benign word error rate (WER)
of 20.85 and 16.62
for the CTC and the decoder heads
respectively.
As has been documented
by previous studies~\cite{jain2018improved,sun2018domain,das2021best},
models trained with MTL also show better benign performance
compared to STL.
For example, we find that training an MTL model
with a decoder and discriminator
($\lambdaTA${}$=$0.8; $\lambdaTC${}$=$0.0)
yields a benign WER of 15.86 on the test set.
For the accent classification task as well,
we observe a reasonable bening accuracy
of $\sim$90\% for the MTL models
with $\lambdaTA < 1.0$.

\begin{figure*}[!ht]
    \centering
    \includegraphics[width=0.98\linewidth]{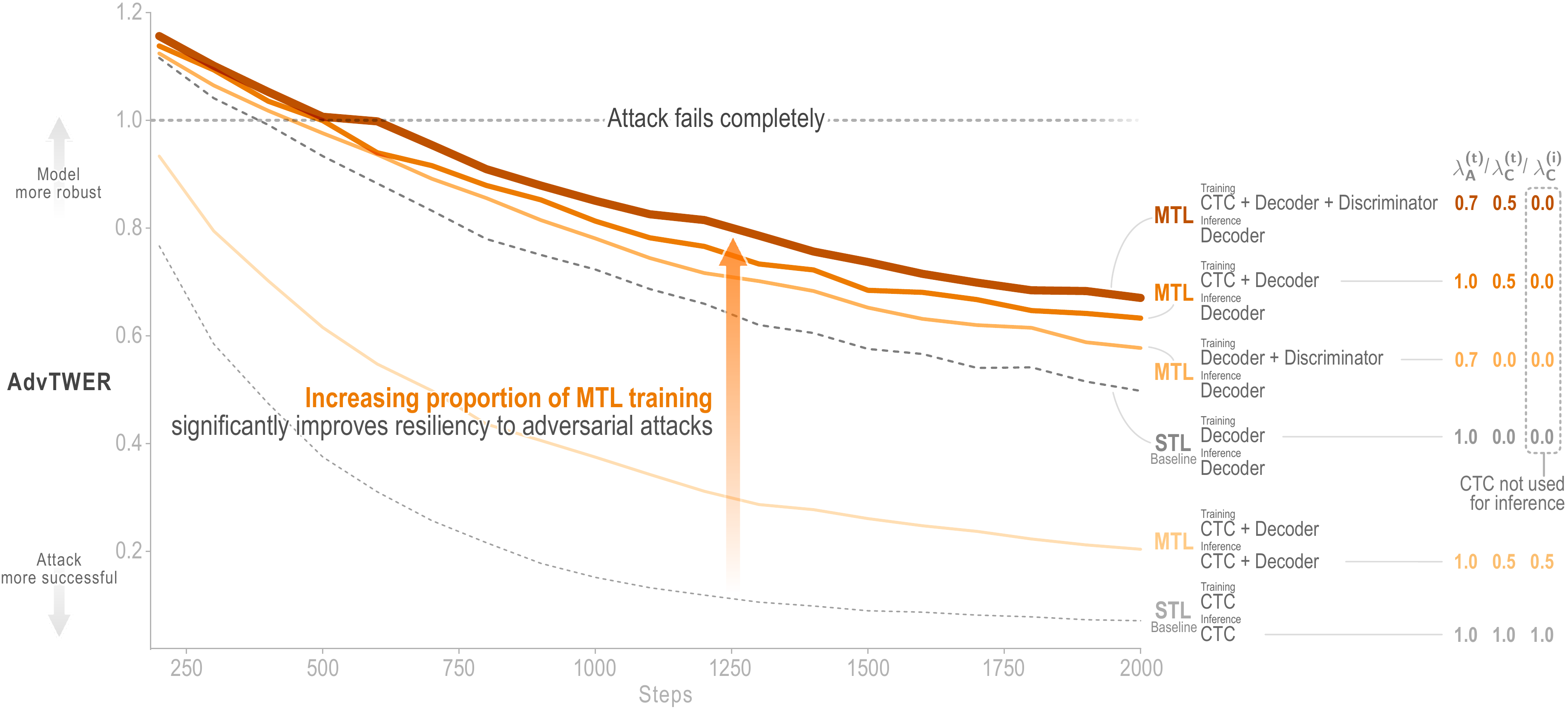}
    \vspace{-0.8em}
    \caption{Adversarial performance for various MTL combinations.
    MTL outperforms STL baselines when CTC is dropped for inference.}
    \label{fig:advtwer_line}
    \vspace{-1em}
\end{figure*}

We now shift our focus
to the adversarial performance
of ASR.
%
%
An overview of the MTL robustness trends
is also depicted in \Cref{fig:advtwer_line},
comparing different MTL combinations
with STL baselines.
We see that
training with MTL
makes it consistently harder
for an attacker to succeed.

\subsection{MTL with CTC and Attention Decoder}
\label{sec:result_sem_sim}
\vspace{-0.4em}

We first study the adversarial performance
of the semantically equivalent task heads
by training jointly
using the CTC and decoder heads.
We train multiple models
by setting $\lambdaTA${}$=$1.0
and modulating $\lambdaTC$
from values $\{0.0, 0.3, 0.5, 0.7, 1.0\}$.
We perform the PGD attack on
each of these models
as decribed in \Cref{sec:setup_adversarial}.
For performing the attack,
we follow two inference modes:
(1) using the CTC head with same inference weights as training,
\textit{i.e.}, $\lambdaIC=\lambdaTC$;
and (2) drop the CTC head for inference,
\textit{i.e.}, $\lambdaIC${}$=$0.0.
It will become clear in the following discussion
why we follow this.
The adversarial performance
for these inference modes
is reported in \Cref{tab:dec_ctc}
using the AdvTWER metric
for various attack steps.

When the trained CTC head is included for inference,
\textit{i.e.}, $\lambdaIC=\lambdaTC$,
we see from \Cref{tab:dec_ctc}
that performing MTL
with a combination of CTC and decoder heads
has no robustness benefit
compared to the STL decoder baseline.
In fact,
the STL CTC baseline itself
is significantly more
vulnerable to the adversarial attacks.
Performing MTL moderately improves
the robustness compared to the STL CTC baseline
but it is still not better than the STL decoder baseline.
This implies that
the attacker is able to
easily influence the CTC head
into hijacking the overall joint prediction.
This can be seen in \Cref{fig:advtwer_line}
with the clear gap between the gray STL baselines
and the MTL model with CTC included in inference
($\lambdaTA${}$=$1.0; $\lambdaTC${}$=${}$\lambdaIC${}$=$0.5).
Decreasing $\lambdaTC$
from 0.7 to 0.3
clearly shows worsening robustness
in \Cref{tab:dec_ctc}.
This means that attacking a less trained CTC head
(with lower $\lambdaTC$)
having a lower impact on the overall prediction
(due to correspondingly lower $\lambdaIC$)
is still able to overpower the
more trained decoder head scores.
This evident vulnerability of the CTC head
may be attributed to
the well studied peaky behavior
of the CTC loss~\cite{liu2018connectionist,zeyer2021does}
that would allow the attacker
to induce overconfident prediction scores
through the CTC head.

Next we study the effect
of performing MTL by training
the CTC head,
but dropping the CTC head during inference,
\textit{i.e.}, $\lambdaIC${}$=$0.0.
This means that the attacker
can no longer manipulate
the CTC head during inference,
but the shared encoder
is still jointly trained
using the CTC loss.
Immediately,
we see the adversarial performance
of the MTL models
beat both the STL baselines
in \Cref{tab:dec_ctc}.
It is consistently harder
for the attacker to
attack the MTL models
across many attack steps.
This is also visualized
in \Cref{fig:advtwer_line}
for
$\lambdaTA${}$=$1.0, $\lambdaTC${}$=$0.5 and $\lambdaIC${}$=$0.0.
%

\subsection{Robust ASR with Semantically Diverse Tasks}
\label{sec:result_sem_div}
\vspace{-0.4em}

We now study
the impact of
performing MTL
with semantically diverse tasks.
As we saw in \Cref{sec:result_sem_sim}
that the CTC head is extremely vulnerable
to adversarial attacks,
we first examine MTL
with the decoder and discriminator combination,
\textit{i.e.}, we set $\lambdaTC${}$=$0.0.
\Cref{tab:dec_dis} shows
the adversarial performance
for training the decoder and discriminator heads
by modulating $\lambdaTA$
from values $\{1.0, 0.9, 0.8, 0.7, 0.6, 0.5\}$.
We see that AdvTWER
for $\lambdaTA${}$=$0.8, 0.7, 0.6
consistently outperforms
the STL decoder baseline.
This implies that
the $\lambdaTA$ should not be too high (less MTL)
or too low (less ASR training)
for optimal MTL robustness.
We can also see this robustness
in \Cref{fig:advtwer_line}
by comparing the STL decoder baseline
with the MTL model for
$\lambdaTA${}$=$0.7 and $\lambdaTC${}$=${}$\lambdaIC${}$=$0.0.
However,
we see that decoder/discriminator MTL
is still not able to beat
CTC/decoder MTL 
when the CTC head is dropped for inference.

\begin{table}[!t]
\setlength{\tabcolsep}{4.8pt}
\centering
\caption{AdvTWER ($\uparrow$ is more robust) with $\lambdaTC${}$=${}$0.0$. Training the decoder with a discriminator shows better robustness.}
\scalebox{0.9}{
\begin{tabular}{l|cccccc} 
\toprule
\multicolumn{1}{r|}{$\lambdaTA \xrightarrow{}$} & 1.0 & 0.9 & 0.8 & 0.7 & 0.6 & 0.5 \\
\midrule
PGD-500 & \textcolor{gray}{93.35} & 89.60 & {\cellcolor{mtlcolor}95.48} & {\cellcolor{mtlcolor}\textbf{97.60}} & {\cellcolor{mtlcolor}97.04} & 90.94 \\
PGD-1000 & \textcolor{gray}{72.31} & 68.67 & {\cellcolor{mtlcolor}73.02} & {\cellcolor{mtlcolor}78.07} & {\cellcolor{mtlcolor}\textbf{78.57}} & 70.25 \\
PGD-1500 & \textcolor{gray}{57.57} & 54.88 & {\cellcolor{mtlcolor}60.88} & {\cellcolor{mtlcolor}\textbf{65.24}} & {\cellcolor{mtlcolor}63.15} & {\cellcolor{mtlcolor}57.88} \\
PGD-2000 & \textcolor{gray}{49.79} & 47.28 & {\cellcolor{mtlcolor}51.23} & {\cellcolor{mtlcolor}\textbf{57.75}} & {\cellcolor{mtlcolor}55.61} & {\cellcolor{mtlcolor}50.43} \\
\bottomrule
\multicolumn{7}{l}{\scriptsize \textcolor{gray}{gray} = STL (Decoder); \hlo{highlighted} = MTL $>$ STL; \textbf{bold} = highest in row}
\end{tabular}
} 
\vspace{-1em}
\label{tab:dec_dis}
\end{table}

Therefore,
we next study the combination of all
three heads
(CTC, decoder and discriminator)
for performing MTL
while dropping the CTC head during inference 
($\lambdaIC${}$=$0.0).
For these experiments,
we first set $\lambdaTA${}$=$0.7
which shows superior adversarial robustness
in \Cref{tab:dec_dis}.
We then modulate $\lambdaTC$
from values $\{0.0, 0.3, 0.5, 0.7\}$.
\Cref{tab:dec_ctc_dis} shows the
adversarial performance results
for this setting.
Similar to \Cref{tab:dec_ctc},
we can see that
increasing MTL training weight
for the CTC head
but dropping the CTC head during inference
improves the overall robustness,
with $\lambdaTC${}$=$0.5
showing the best robustness
of all MTL models.
Hence,
combining all three heads
and performing MTL 
with semantically diverse tasks
makes the model most resilient
to the adversarial attacks
consistently across multiple attack steps.
This can also be clearly seen
in \Cref{fig:advtwer_line}
where the MTL model
corresponding to
$\lambdaTA${}$=$0.7, $\lambdaTC${}$=$0.5 and $\lambdaIC${}$=$0.0
outperforms all other MTL combinations and baselines.

\begin{table}[!t]
\setlength{\tabcolsep}{9pt}
\centering
\caption{AdvTWER ($\uparrow$ is more robust) with $\lambdaTA${}$=${}$0.7$, $\lambdaIC${}$=${}$0.0$. 
Training all heads combined
shows most superior robustness.}
\scalebox{0.9}{
\begin{tabular}{l|cccc} 
\toprule
\multicolumn{1}{r|}{$\lambdaTC \xrightarrow{}$} & 0.0 & 0.3 & 0.5 & 0.7 \\ 
\midrule
PGD-500 & \textcolor{gray}{97.60} & 97.02 & {\cellcolor{mtlcolor}100.65} & {\cellcolor{mtlcolor}\textbf{101.04}} \\
PGD-1000 & \textcolor{gray}{78.07} & 76.83 & {\cellcolor{mtlcolor}\textbf{85.06}} & {\cellcolor{mtlcolor}83.69} \\
PGD-1500 & \textcolor{gray}{65.24} & 62.75 & {\cellcolor{mtlcolor}\textbf{73.70}} & {\cellcolor{mtlcolor}70.67} \\
PGD-2000 & \textcolor{gray}{57.75} & 56.57 & {\cellcolor{mtlcolor}\textbf{67.04}} & {\cellcolor{mtlcolor}65.79} \\
\bottomrule
\multicolumn{5}{l}{\scriptsize \textcolor{gray}{gray} = Baseline; \hlo{highlighted} = MTL $>$ Baseline; \textbf{bold} = highest in row}
\end{tabular}
} 
\vspace{-1em}
\label{tab:dec_ctc_dis}
\end{table}

\section{Conclusion}

In this work,
we study the impact of
multi-task learning (MTL)
on the adversarial robustness
of ASR models.
We perform extensive experimentation
with multiple training and inference hyperparameters
as well as wide-ranging adversarial settings.
Our thorough empirical testing
reveals that
performing MTL 
with semantically diverse tasks
consistently makes it harder
for an attacker to succeed
across several attack steps.
We also expose the extreme vulnerability
of the CTC loss,
and discuss related pitfalls and remedies
for using the CTC head during training with MTL.
In future work,
we aim to investigate
regularization methods
that can reduce the peaky behavior
of CTC so as to
induce more adversarially robust
hybrid inference.

\bibliographystyle{IEEEtran}
\bibliography{references}

\end{document}